# Search for a Dark Matter component

Alessandro Giordano[1], George N. Izmailov[2‡], Roberto De Luca[1†], Andrej M. Tskovrebov[3], Larisa N. Zherikhina[3], Vladimir A. Ryabov[3]

[1] *Department of Physics, University of Salerno, I-84084 Fisciano (SA), Italy*
[2] *Moscow Aviation Institute (National Research University), Moscow, Russia*
[3] *P. N. Lebedev Physical Institute, Moscow, Russia*

E-mail: [†] rdeluca@unisa.it
        [‡] izmailov@mai.ru

## Abstract

In this paper, after reviewing some of the most important concepts about Dark Matter (DM) and methods of its registration, in particular by using SQUIDs, we focus on two main problems. First, the possible mechanism of magnetic moment origin for DM particles, in the form of neutralino, is discussed: the presence of a magnetic moment means the existence of a new kind of interaction, whose corresponding cross section is estimated. Second, a simple uniform model for DM and Dark Energy (DE) is proposed. Two types of devices based on SQUID, in particular the SQUID-paramagnetic absorber and the SQUID-magnetostrictor systems, both suitable for investigations of above problems, are considered.

## 1. Introduction

The enigma of Dark Matter (i.e. non-luminous and non-light absorbing matter) is one of the major open problems of modern science. Swiss astronomer Zwicky was the first to suggest, in 1933, the existence of Dark Matter (DM) on the basis of observation of the velocity dispersion of eight galaxies in the Coma Cluster [1]. The discovery of DM played about the same role, in cosmology, as the discovery of radioactivity phenomena by A. A. Becquerel, at the very end of the 19th century, had played in nuclear physics. Very soon after the first registrations of nuclear radiation it became clear that the well-known electromagnetic forces (actually described in the frame of Special Theory of Relativity) appeared to be much smaller than the forces of nuclear nature. The DM existence, according to modern ideas, demonstrates that the effect of gravitational curvature in Universe, described within the General Theory of Relativity, is negligible in comparison with interactions manifested by DM. The nature of this new effect, strong with respect to the common relativistic curvature of space, is yet obscure. The only thing we can say about DM is that its elementary particles must be massive and have no electric charge, but since DM does not radiate light and can gravitationally interact with other celestial bodies.

In succeeding years a great deal of convincing evidences of the DM existence was obtained at various scales (see, for example, reviews [2-4]). A striking picture of Universe arises as the result of investigations pursued since the beginning of this century. It consists of 2/3 of some repulsive cosmological component (Dark Energy) and 1/3 of matter, 85% of which is DM, whereas only 5% of the Universe content is accounted for ordinary (baryon) matter [5-7]. The role of repulsive (antigravity) component is intimately related to the universal cosmological constant problem, initially appeared in the theory of general relativity (the $\Lambda$ term in Einstein's equations) and currently discussed in the Superstring Theory framework [8].

Modern theoretical models offer a broad assortment of particles which could constitute DM (see, for example [2-4]). The most popular candidates are particles generally called WIMPs (Weakly Interacting Massive Particles) with masses from a few tens of GeV to several TeV. (In the context of DM, hypothetical light particles (ALPs) with masses much smaller than 1 GeV, forming a hot component of DM, are also currently considered. However, only the registration of cold nonrelativistic DM components will be discussed below. See also Figure 3).

In the Minimal Supersymmetric Standard Model (MSSM), the lightest neutralinos (linear combinations of four neutral fermions: Wino, Bino, and a pair of Higgsinos) are appropriate candidates.

All experimental studies on the search for DM particles can be conditionally divided into three main areas [4]: 1) experiments on colliders; 2) indirect registration of Dark Matter particles by their annihilation products in cosmic rays; 3) direct detection of Dark Matter particles of cosmic origin.

Accelerator experiments (Tevatron, LHC), may give certain results only on the basis of full kinematic analysis of visible products of *p-p* interaction, allowing to recover the value of the energy-impulse "spent" at the birth of the unknown DM particles. It is estimated that only a small part of the total energy of *p-p* collisions is spent for the creation of supersymmetric particles, which limits the effectiveness of the experiment. Thus the origin of gluino and squark should demand about 10 % of the kinetic energy of the protons, so the eventual formation of a 100 GeV WIMP is supposed to be implemented with energy not less than 2 TeV.

Indirect registration of DM particles, by their annihilation products in cosmic rays, requires the detection of TeV gamma rays. However, as it is well known [9], such a quantum creates in the Earth's atmosphere a wide (a few kilometers) electromagnetic air shower of secondary particles that significantly complicates the determination of the total energy of the original photon. Among the projects for registration of 1÷15TeV gamma rays in space-based experiments, it should be noted the project GAMMA-400, developed in P. N. Lebedev Institute [10, 11], as one of the most competitive (energy resolution of 1%, angle resolution 0.01°). This gamma-ray telescope is a stack of silicon strips and scintillation plates, and a TeV quantum, passing through it, practically loses all its energy. The system of photomultipliers allows not only to determine the initial energy of the quantum, by summing photo responses, but also to identify the point of conversion and the direction of the incident photon with the help of fiber-optic cabling.

The experiments for direct registration of DM particles of space origin are carried out in laboratories with deep depression of cosmic background (the radiation weakening in the tunnel of Gran Sasso is characterized by the water equivalent of about 3600 m). By comparing the characteristic energy spectrum of recoil nuclei with the corresponding spectrum of known weakly interacting particles (for instance, neutrinos), also able to go through very thick defense shields, one can reach the same conclusion about detection of WIMPs. For the registration of recoil nucleus energy, a wide range of detectors is used in these experiments [4]: ionization, scintillation, phonon, Josephson junction, heat, based on the first order phase transition (such as bubble chambers or superheated superconducting microgranules) and so on. In the DAMA project [12], for instance, there are 9 scintillating amplitude detectors (crystals of NaI (TI), each one of 9.7 kg) with energy resolution of about 2 keV. This project should be especially mentioned. In fact, during its seven-year-long observation period, it really fixed for the first (and in practice for the last) time the annual cycles of decreasing and increasing character of the registration rate of events. In practice, it provided clear comprehension of the coinciding/anticoinciding (June/December) direction of the velocity vector of galactic streams of Dark Matter particles with the travel line through solar system. There is now a new version of this experiment, called DAMA LIBRA, using 25 scintillators of the same kind [13].

The general development trend of modern methods for Dark Matter particles direct detection is turned to the wide introduction of cryogenics. The evident lowering of thermal noise, along with the drifting of input stages of used electronics, permits in numerous cases to engage fundamentally new effects (the low-temperature phase transitions of the first order, SQUIDs, Josephson arrays [14,15], multi-barrier Josephson junctions [16], etc.) in order to increase the sensitivity of the registration process itself [4]. Recording circuitries, where the amplitude measurements of detector response are registered by means of SQUIDs [17-19], based on Josephson effect [20], have had a wide spread. The sensitivity of a modern commercial DC-SQUID (without superconducting flux transformer) reaches the level of $10^{-6} \div 10^{-7}$ $\Phi_0/\sqrt{Hz}$ (here we have the flux quantum $\Phi_0=\pi\hbar/e\approx2,07\times10^{-15}$Wb). The sensitivity of SQUIDs has permitted to apply these devices for taking such ultra-precise measurements as gravitational wave detection [21, 22] or contactless examination of bio magnetic brain activity [21, 23]. According to the physical principles of its work, the quantum interferometer measures the magnetic flux. By using

Stokes flux theorem, we can determine the interference phase difference in superconducting circuit, where Josephson junctions are included [17,18]. However, in experiments for DM particles search, these devices are usually used merely as low-frequency picovoltmeters, registering the response of cryogenic thermoresistors. In this way, in the two-segment detectors (the project of CRESST-II [24]), the coincidence of light and thermal responses of absorber (300g of $CaWO_4$) on recoil nucleus are fixed. Two identical vanadic thermoresistors, being at a temperature near the superconducting transition, together with two DC-SQUIDs, are used as scintillating and heat recording channels of Dark Matter particles. However, alike schemes of quantum interferometer utilization, when the signal is converted according to the chain $I\delta R \rightarrow \delta U \rightarrow \delta i = \delta U/r \rightarrow \delta \Phi = L\delta i$, i. e. at first it turns into the variance of magnetic flux taken by the interferometer, and then it is directly applied at the SQUID entry, happen to be inefficient due to unavoidable losses in the conversions chain. However, two research groups [25, 26] have proposed schemes, that did not need any conversion of signal. Actually the heat response, arising in absorbers due to the interaction of a particle with matter, was transmitted directly on the signal input. In ref. [25] the possibility to measure the magnetic response of paramagnetic thermometer, by means of DC-SQUID, was tested. The signal appears here due to the dependence of thermal sensor magnetic susceptibility according to Curie-Weiss law. In this case the sensor should be magnetized by a small magnetic field (~10 mT). We have proposed [26] another type of scheme, where the magnetic response, registered by a SQUID, corresponds to an enhancement of the spin system entropy rather than to the change of paramagnetic absorber temperature. The operating principle of the experiment is the following. At the beginning the paramagnetic absorber is self-cooled during the process of adiabatic demagnetization [27]. After the external magnetic field is lowered down to zero, the SQUID measures the reduction of residual magnetization of the paramagnetic absorber. The latter step corresponds to the measurement of the entropy increase due to the release of energy caused by the interaction between a particle and the paramagnetic absorber. Various modes of operation of the SQUID-paramagnetic absorber system are discussed in details in the following works. In ref. [28] direct measurement of the entropy growth using the method of adiabatic demagnetization is considered. In ref. [29] the measurement sensitivity is increased by replacing atom paramagnetism by nuclear one, with cooling produced by a dissolution refrigerator $He^3$ - $He^4$. In ref. [30] a dual-channel mode is adopted to eliminate lepton processes. In ref. [31] an estimate of the sensitivity in the case of strong fields saturation using asymptotic methods of statistical mechanics is performed. Finally, in ref. [32] resonance registration of THz radiation with a wavelength of about 10 mm is illustrated.

It should be noted that, among all thermal methods of Dark Matter particles registration, the magnetic ones have two essential advantages. In fact, in order to increase the probability of registration of elementary particles, which weakly interact with matter (small cross-section of interaction, approximately one event per kg per day), it is necessary to enlarge the mass of the absorber. As a consequence, the mass of real detectors of Dark Matter particles evolved from the initial value of 100 g to 100 kg [33,34] in a short time period [4]. In common (nonmagnetic) thermal detectors the enhancement of the absorber mass automatically leads to the enhancement of its thermal capacity and hence to a sufficient reduction of the thermal response. On the contrary, in cryomagnetic systems the enhancement of heat capacity is compensated by the simultaneous increase of the total number of spin-carrying particles contributing to the system's magnetic response. Another useful property of magnetic thermal detectors is connected with the growth of the magnetic part of heat capacity as the temperature decreases: $c_m \sim T^{-2}$. This fact seems paradoxical at first glance, since it may appear inconsistent with the third law of thermodynamics. However, this dependence is true until either the total ordering due to ferromagnetic transition or the local ordering due to casual residual fields take place. In the case a little magnetic sensor with a large magnetic heat capacity is attached to a big nonmagnetic absorber, characterized by a small heat capacity, the whole energy released in the latter can be gathered in the former device.

In the next section we consider the possible magnetic interaction of DM particles with common matter. We shall deal with a new kind of interaction, different from the conventional WIMP-nucleus scattering. In fact, this interaction is an action at distance, and its analysis is very interesting in trying to open new horizons on DM hunt. One of the most recent experiment, Xenon 100 [35], performed in the underground laboratory of Gran Sasso, has excluded hidden WIMP-electrons interactions. In ref. [36] the sharp difference between values of cross sections for spin dependent (SD) interactions and spin independent (SI) ones (the former are nine orders of magnitude bigger than the latter), is remarked. It is noted that an adequate model of such SD interaction for DM detection is still needed.

In the third section we propose a second type of experimental device, the SQUID-magnetostrictor system, in order to register DM fluxes. In addition, a model of unified DE-DM providing definition of this flux is given. Conclusions are finally drawn in the last section.

## 2. Cross section estimate of the magnetic interaction of DM particles for SQUID registration

One of the most realistic hypothetical candidate for cold DM is the lightest supersymmetric particle, the so called neutralino, whose wave function is a linear combination of fermionic super-partners of photon, of *W*-neutral boson and of Higgs boson. This wave function can be thus denoted as $\aleph = N_{11}\hat{B} + N_{12}\hat{W}_3 + N_{13}\hat{H}_1^0 + N_{14}\hat{H}_2^0$, where $N_{11}$, $N_{12}$, $N_{13}$, $N_{14}$ are some opportune constants (being the lightest supersymmetric particle, neutralino should be stable). Of course, being "neutral in all respects", neutralino has no electric charge; however, electro-neutral elementary particles can possess a magnetic moment. In general, a magnetic moment might occur for two main reasons: first, because of the reversible virtual transformation of the original "non-magnetic" particle (in its ground state) to one particle of the multiplet partners with an electric charge (SU(2) baryons with isospin ½: (a neutron *n* into a proton *p*)); second, because of the existence of a cloud of virtual charged quanta of the interaction field, involving "naked nonmagnetic" particles. According to these modern concepts, the neutron magnetic moment is (approximately) analogously formed.

Similarly, a very weak magnetic moment of the neutrino ($\mu_\nu \approx 10^{-13} \mu_B$) should occur [37] due to the electroweak processes illustrated by Feynman diagrams represented in Figure 1.

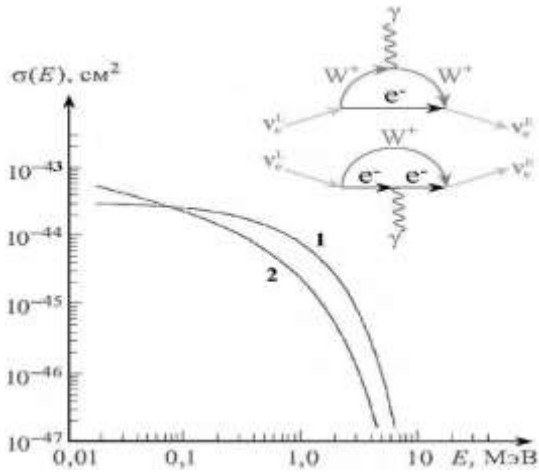

**Fig 1.** Cross section of neutrino scattering on electron: **1** - weak interaction (the Weinberg's angle agrees with $sin^2\theta_W=0.23$); **2** - magnetic interaction ($\mu_\nu=10^{-13}\mu_B$) [36]. In the inset, Feynman diagrams illustrating the creation of anomalous magnetic moment of Dirac (massive) neutrino $\nu_{(D)}$ are given.

In the framework of Weinberg-Salam electroweak interactions (Standard Model), electron neutrino $\nu_e$ has some non-zero probability to decay into an electron and a $W^+$ boson and then through a time interval $\Delta t \approx \hbar/(m_W c^2)$ these virtual particles annihilate, turning into another helicity neutrino. During the short ($\approx 2\times 10^{-27}$ sec) existence of electrically charged particles $e^-$ and $W^+$, they can interact with an external electromagnetic field, which is symbolized in the diagram by a photon $\gamma$. Therefore, the part of the radiative corrections, which determines the energy shift, is interpreted as the interaction energy of the neutrino magnetic moment with the magnetic field. On the other side, some astrophysical estimates [38, 39] lead to the hypothesis about a significantly larger neutrino moment value than the one given by the Standard Model. In the 90's, the search for such anomalous magnetic moment of neutrinos was engaged, in particular, by B. S. Neganov (the one who invented the dilution cryostat $He^3$-$He^4$ for obtaining temperatures below 100 mK without magnetic field) at JINR in Dubna. In his experiments [40] an attempt to observe the growth of the interaction cross section e-/ν predicted for the "magnetic" neutrinos in the region of small energy was undertaken. A low-temperature calorimetric detector and a $H^3$ source of neutrinos were used.

Similar assumptions can be considered about the presence of a magnetic moment for DM particles, also if they are beyond the Standard Model. One of the channels [41] is the reversible annihilation of the neutralino into a pair of electrically charged gauge bosons W-type.

A diagram illustration of the process 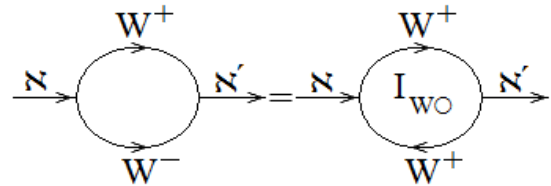

shows that two branches of virtual oppositely charged $W^-$ and $W^+$ bosons form a ring current $I_{WO}$, whose corresponding boson loop has the uniform charge (and effective area of $S_{WO}$). In fact, we may notice that the two diagrams are equivalent, in the sense that the motion of the $W^-$ boson in one direction is equivalent to the motion of the $W^+$ boson in the opposite direction. In this way, the corresponding ring electric current of $W^+$ bosons is formed inside the considered loop. We can express the magnetic moment of the boson loop as $S_{WO} I_{WO}$, which is identified as the magnetic moment of a neutralino. (It should be noted that the specificity of magnetic interaction of DM particles with common matter is in its "tangential character", as opposed to the conventional "nuclear head on" type).

In order to obtain an analytic expression for it, we may consider the analytic expressions for the effective area and the ring electric current. The effective area can be represented through the square value of the Compton wavelength $S_{WO} \approx \hbar^2/(m_W c)^2$, which is the minimum possible value, as the speed of light $c$ is the maximum value for velocities. The loop electric current can be estimated as $I_{WO} \approx e/\tau_W \approx e m_W c^2/\hbar$, where $\tau_W$ is the minimum value for the time interval, according to Heisenberg's uncertainty principle, and so $I_{WO}$ is the maximum possible value for the considered loop of electric current. So the final expression $S_{WO} I_{WO} \approx e\hbar/m_W = \mu_W$ is an adequate analytic representation for the neutralino magnetic moment.

This analytic expression coincides with the structure of the standard formula of Bohr magneton $\mu_B$ and differs from the latter by replacing of the electron mass $m_e$ with $m_W \approx 1,6 \times 10^5 m_e$. Accordingly, $\mu_W$ is approximately 5 orders of magnitude smaller than $\mu_B$. This means that the magnetic interaction energy $E_N = \mu_W B_{orb}$ of $\mu_N$ with magnetic induction $B_{orb} \approx 10$ T (typical value of the field for spin-orbit effects) is about $4 \cdot 10^{-9} eV$. The probability of interaction between an atom in the absorber, whose atomic orbital current induces a magnetic field, with the magnetic moment of the boson loop, occurring during the reversible *decay* of neutralino, can be estimated by squaring the corresponding variation $\delta\psi_{Wo}$ of the amplitude of the unperturbed boson loop, defined according to the following expression:

$$\psi_{W\circ} = \psi_{W\circ}^{(0)} + \delta\psi_{W\circ} = \psi_{|W\circ\rangle}^{(0)} + \sum_{\forall |W\circ\rangle'} \frac{\mu_W B_{orb}}{E_{|0\rangle} - E_{|W\circ\rangle'}} \psi_{|W\circ\rangle'}^{(0)}$$ [42]. Thus, the required probability

for a typical value of the energy, eventually lost by neutralino, $\delta E_\aleph \approx 40$ eV (energy quantity transferred to the absorber in case of a reliable registration), is estimated as $(\mu_W B_{orb}/\delta E_\aleph)^2 \approx 10^{-20}$. Therefore, the probability sought for a typical value of energy, eventually lost by a neutralino, and coinciding with $\delta E_\aleph \approx 40$ eV (that is roughly transferred to the absorber in a reliable registration), is estimated by squaring the ratio between the corresponding variation and the amplitude of the unperturbed boson loop, so we get: $(\mu_W B_{orb}/\delta E_\aleph)^2 \approx 10^{-20}$. Considering a linear chain of $10^{20}$ absorber atoms, and adding the probabilities of magnetic interaction with all atoms, as all the events of interaction are independent from one another, and also characterized by the same probability, we get the level of confidence $\sum_{1}^{10^{20}} 10^{-20} \cong 1$.

According to this expression, we have almost the certainty that a magnetic interaction event occurs between the neutralino and one of the atoms of the considered linear chain, having a length of $a \cdot 10^{20}$ where $a$ is the lattice constant, which assumes a value of about 0.3 nm. Let us "build" a hypothetical absorber with a large number of such chains and let its square surface area be $S_0$. Then the cross section of the magnetic interaction neutralino-absorber will be $\sigma_{\aleph \leftrightarrow Borb} \approx \sigma_{W\circ \leftrightarrow Borb} \approx S_0/N_A = S_0/(S_0 \times a \times 10^{20} \times n_A) \approx 10^{-20}/(a \times n_A) \approx 10^{-35}$ cm$^2$, where $n_A = a^{-3} \approx 3 \times 10^{22}$ cm$^{-3}$ is the concentration of atoms in the absorber. Moreover, this implies that the less is the registered energy, the more such events should occur, and hence the higher is the estimated cross section. Estimated «magnetic cross section» at the level of $10^{-35}$ cm$^2$ happens to be noticeably higher than typical values of level of $10^{-44}$ cm$^2$ for the conventional interaction WIMP-nucleus. If it is true, we can consider some features of optimal experimental design for search of DM particles such as the neutralino. A calorimeter with a possibly low energy detection threshold (not higher than $\delta E \approx 40$ eV) is required. In addition, a solid-state absorber made of atoms with strong spin-orbital effect, indicating the presence of a large (not lower than $B_{orb} \approx 10$ T) orbital magnetism, is needed. The only candidate for the role of calorimetric detector with energy threshold of the order of $\delta E \approx 40$ eV is the SQUID-paramagnetic absorber system [25, 26]. This cryogenic system (Fig. 2) consists of a paramagnetic absorber, demagnetizing as a consequence of the heat transferred by the energy $\delta E$ of the detected radiation, and of a quantum interferometer, measuring the corresponding decrease of the magnetic moment $\delta m_{ads}$ of the absorber.

At sufficiently low temperatures ($T \approx 1$ K), the contribution of the atomic paramagnetism [27] to the heat capacity prevails on the phonon contribution, so the following relation holds: $\delta E \approx B \delta m_{ads}$ (where $B$ depends on the mode of operation of the system, and it may be either the induction of the external magnetizing field [25] or the residual paramagnetic field [26]). The magnetic flux variation, directly registered by the SQUID, is $\delta\Phi \approx \mu_0 \delta m_{ads}/h \approx \mu_0 \delta E/(hB)$, where $h$ is the absorber length (the height of the paramagnetic cylinder), and $\mu_0 = 4\pi \times 10^{-7}$ H/m.

The Superconducting Quantum Interference Device (SQUID) [17-19]), due to the sensitivity of its Josephson junctions, to the difference between the Cooper condensates quantum phases [20] and to the detected magnetic field flux inside the superconducting ring, fixes flux variations as a fraction of the basic period, that is the flux quantum $\Phi_0 = \pi\hbar/e \approx 2.07 \times 10^{-15}$ Wb (which corresponds to a phase change $\delta\varphi = 2\pi$). At the same time a good, but not record sensitivity, of the modern interferometer is considered to attain the value $\delta\Phi \approx 10^{-6} \Phi_0/\sqrt{Hz}$. This value corresponds to the energy resolution $\delta E \approx hB\delta\Phi/\mu_0 \approx 2 \times 10^{-18}$ J/Hz $\approx 15$ eV/Hz, if $h \approx 0.1$ m and $B \approx 0.01$ T, that makes it possible to fix $\delta E \approx 40$ eV with maximal frequency nearly 10 events per second. However, real conditions of experiment on the Earth correspond to the density of DM particle flux at the level of no more than 200 km/s × 1500 particles/m$^3$ = $3 \times 10^8$ s$^{-1}$m$^{-2}$ with respect to the absorber. If we use as an absorber a paramagnetic material with strong atomic orbital magnetism with volume $h \times S \approx 0.1$m × 0.01m$^2$, it would contain approximately 0.15 kmole $\approx 10^{26}$ atoms. The cross section of interaction is then $\sigma \approx 10^{-35}$ cm$^2$, ensuring a maximum registration rate of $3 \times 10^{-5}$ events/s $\approx$ 4 events/day. Therefore, the margin of recording rate of about 6 orders of magnitude ($10/3 \times 10^{-5}$) can be used to compensate the loss of sensitivity of the system, associated with a low transmission coefficient of the superconducting flux transformer (K<1, depending on the design [19]). This transformer provides communication between the macroscopic working body of the absorber with the microscopic phase-sensitive ring of the SQUID, where the Josephson junctions are allocated (such compensation is possible up to the level of K$\approx(3 \times 10^{-5}/10)^{1/2} \approx 0.0017$).

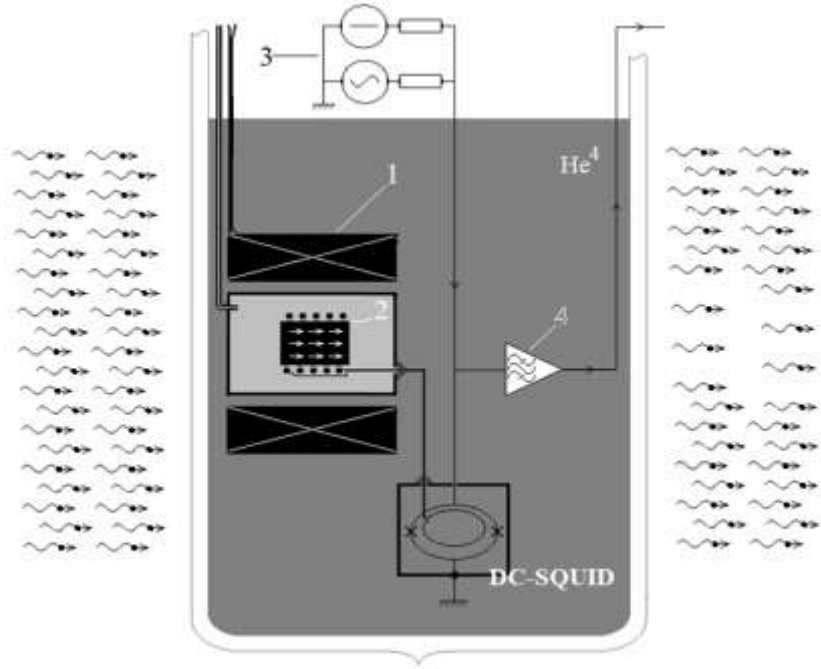

**Fig. 2.** Scheme of the SQUID-paramagnetic absorber system: 1) – the superconducting solenoid magnetization; 2) the paramagnetic absorber; 3) current generators; 4) narrow–band low-frequency amplifier.

## 3. Non-corpuscular "ether wind" and possible registration of its pressure by the SQUID-magnetostrictor system

In this section, we consider a new approach for DM description, based on the possibility that DM and Dark Energy (DE), can be considered as two different aspects of the same cosmological essence, named "Dark Substance" (DE - hypothetical pervasive substance, which can be responsible for the additional relative acceleration in the Hubble law of galaxies recession). According to this model DE, with its density of about 300 TeV/m$^3$, represents the unperturbed state of "Dark Substance", while its swings or perturbations play the role of elementary DM particles. These particles will be stable if all their decay channels into any combination of other particles are blocked, or also, in our case, if their potential will have local minima, i.e., local traps providing metastable excited states. The Hamiltonian with metastable traps can be represented, for example, as follows: $\mathcal{H} = \frac{1}{2}\left(\frac{\partial \varsigma}{\partial x}\right)^2 + \frac{1}{2c^2}\left(\frac{\partial \varsigma}{\partial t}\right)^2 + \alpha \varsigma^* \varsigma - \beta \cos \gamma \sqrt{\varsigma^* \varsigma}$.

The nonlinear wave equation, corresponding to this Hamiltonian, and describing the dynamics of perturbations to Dark Substance will be like the "quasi-sine-Gordon" equation $\frac{\partial^2 \varsigma}{\partial x^2} - \frac{1}{c^2}\frac{\partial^2 \varsigma}{\partial t^2} + \alpha \varsigma^* + \frac{1}{2}\beta\gamma \sqrt{\varsigma^*/\varsigma} \sin \gamma \sqrt{\varsigma^* \varsigma} = 0$. Moreover, the nonlinear potential

$\Pi(\varsigma)=\alpha\varsigma^*\varsigma-\beta\cos\gamma\sqrt{\varsigma^*\varsigma}$ appearing here, has the analytic structure very similar to the "parabolic washboard potential" [17] used to describe metastable states in the superconducting ring of a SQUID with one Josephson junction, namely: $\Pi(\varsigma)=\alpha|\varsigma|^2-\beta\cos\gamma|\varsigma| \Leftrightarrow E(\Phi)=\frac{1}{2L}\Phi^2-I_C\Phi_o\cos\frac{2\pi}{\Phi_o}\Phi$, where $\Phi$ is the magnetic flux threading the superconducting ring and $L$ is the inductance of the loop itself. We represent the profile of the washboard potential in figure 3. Some comments about this figure are needed. For small $\varsigma$ (when the disturbance has not yet reached the first trap) swings of the Dark Substance have a quasi-harmonic character, and their quanta will have a mass $m=\frac{\hbar}{c}\sqrt{2\alpha+\beta\gamma^2}$. However, such particles may be unstable, and the rate of decay from classical positions will correspond to the viscosity of the Dark Substance [43, 44]. At large amplitudes of $\varsigma$, the disturbance at some moment will "catch" the lowest energy trap. The trajectory of the oscillation $\varsigma$ will now be a circle in the plane $\{\varsigma, \varsigma^*\}$, corresponding to a local minimum of potential. The rotation around the circumference of a local minimum is similar to the mechanism of occurrence of massless Goldstone bosons in Weinberg-Salam's model. However, in this example, the mass of excitations ("zero energy") is determined by the height of the bottom of the trap with respect to the main vacuum state $\varsigma=0$ $\varsigma^*=0$, and will be non-zero. The stability of such excitations, playing (in this example) the role of DM particles, is guaranteed by the height of the wall of the potential well, occurring in the vicinity of the local potential minimum.

Therefore, the search of DM particles, as stable moving excitations of Dark Substance, may be intimately connected with the research of the action of the DE non-corpuscular flux on the ordinary matter. By knowing that the free mean path is connected to the cross section of interaction by the relation $\ell^* \approx 1/(\sigma n_A)$, we may say that DE transfers to a slab of material, consisting of ordinary atoms of concentration $n_A$, with area $S$ and "maximum depth" $\ell$, a momentum $q=\ell^* S \rho_{DE}/\upsilon$, where $\upsilon$ is the speed of DM particles relative to the substance. In this way, Dark Substance exerts the pressure $p^*_{DE}=F/S=(\ell^* S\rho_{DE}/\upsilon)/(\ell^*/\upsilon)/S=\rho_{DE}$ on the slab. Thus, the effective pressure drop across the length $\ell$ is estimated as $p_{DE}=p^*_{DE}\ell/\ell^*=\rho_{DE}\ell\sigma n_A$. In accordance with the generally accepted value of the average density of Dark Energy $\rho_{DE} \approx 300$ TeV/m$^3$, taking into account the above-obtained "optimistic" estimation of the interaction cross section $\sigma\approx10^{-35}$cm$^2$, the pressure drop across a one-meter barrier, having a concentration of atoms $n_A\approx3\times10^{22}$cm$^{-3}$, will be of the order of $p_{DM}(\ell=1м)\approx7\times10^{-16}$Pa.

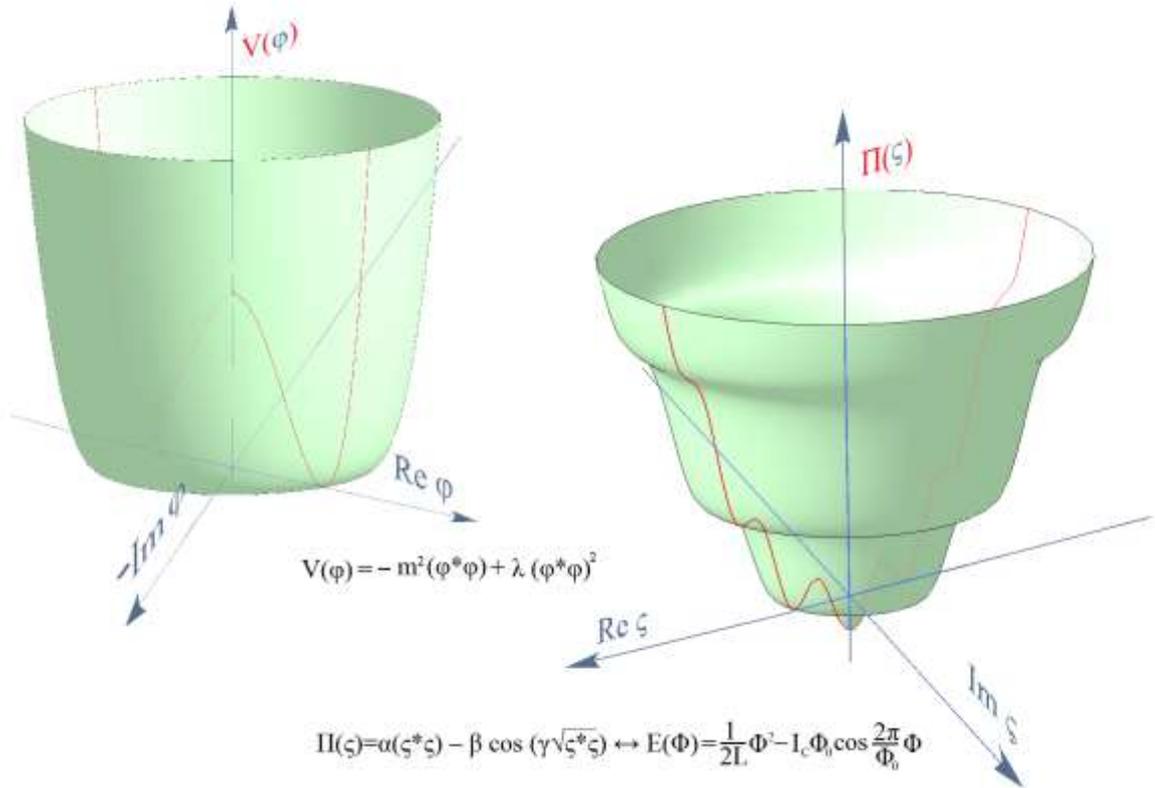

**Fig 3.** On the right, the washboard potential, characterized by local minimum positions, which are the metastable states of DM particles, is shown. The lower local minimum is associated with a light (or also hot) component of DM, and the top one with a heavy (or also cold) component of DM. The absolute minimum lies in the region of negative energies and is associated with antigravitational properties of DE. On the left (for comparison) the potential used in models of spontaneous symmetry breaking is displayed.

In order to register this pressure, a dynamometer capable to fix the strength of $10^{-16}$ N at the end of the cylinder, with dimensions $\ell \times S \approx 1\text{m} \times 0.15\text{m}^2$, is required. Apparently, the only candidate for the role of the super-high-sensitivity dynamometer is the SQUID-magnetostrictor system [22, 45,46] which has been previously supposed to be used in projects for the detection of gravitational waves, etc. (Figure 4).

Ultra-high sensitivity is achieved by means of this system. In fact, high strain-gauge effectiveness of the sensor can be achieved, since it operates on the principle of the reverse magnetostrictive effect, generated, in its turn, by collective quantum solid-state effects [46]. On the other hand, the high ("quantum scale" level) sensitivity of SQUID systems, used as measuring instruments, allows accurate registration of events.

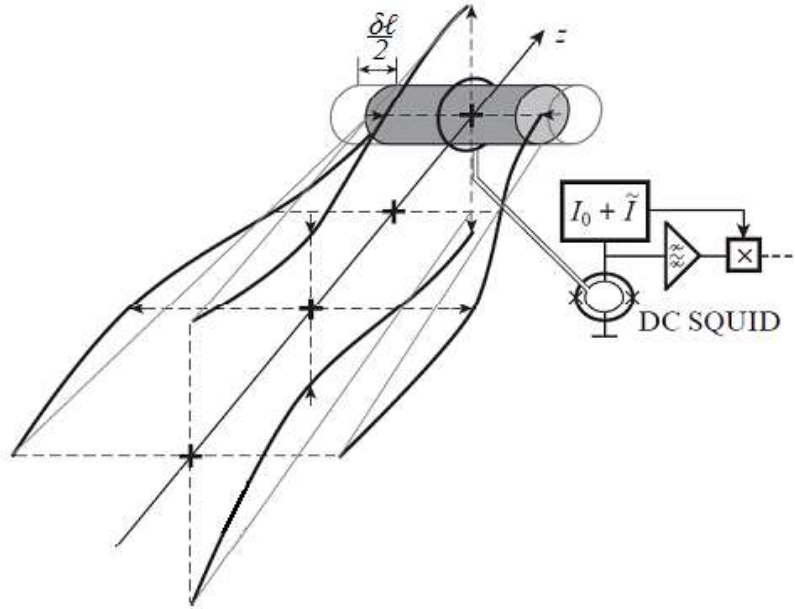

**Fig 4.** Schematic view of the SQUID-magnetostrictor system for the detection of gravitational waves (the magnetostrictive cylinder is represented in grey) [46].

The physical quantity describing the reverse magnetostrictive effect (discovered by Emilio Villari in 1865) in a particular material is the ratio of the internal magnetic induction field to the growth of its outside pressure, i.e., $\Lambda^{-1} = \frac{\partial B}{\partial P}$. For example, an alloy made up of 54% Pt and 46% Fe, with $\mu \approx 14000$, will have $\Lambda^{(-1)} \approx 10^{-4}$ T / Pa (which is basically not a record value). Thus the magnetic response, measured by the SQUID, is related to the force action $\delta F$ by this expression $\delta\Phi = \Lambda^{(-1)} \delta F$. Accordingly, the capability to register the pressure of non-corpuscular Dark Matter flow, estimated above for $\sigma \approx 10^{-35} cm^2$ at $\delta F \approx 10^{-16}$ N, requires an "absolute" (not reduced to the time of the signal accumulation) SQUID sensitivity for the magnetic flux of the order of magnitude of $10^{-20}$ Wb $\approx 5 \times 10^{-6} \Phi_0$. The actual value of a good DC-SQUID is of about $\delta\Phi \approx 10^{-6} \Phi_0/\sqrt{Hz}$, which provides the desired sensitivity with a margin of approximately 2 orders of magnitude (at least) due to the possibility of a 3-hour signal accumulation.

## 4. Conclusions

In this paper, starting from an introduction about DM and its cosmological properties, we have focused our attention on the possible creation of neutralino magnetic moment, the estimation of its magnetic interaction cross section, and a brief description of a model unifying DE and DM. In particular, we have found magnetic moments for neutralino to be 5 orders of magnitude smaller than the one for electrons, and a magnetic interaction cross section that is 9 orders of magnitude larger than the one corresponding to the conventional interaction WIMP-nucleus. The specificity of this magnetic-type interaction is in its "tangential character", as it is an interaction at distance with the magnetic field induced by the orbital motion of atomic electrons. So, there is a remarkable difference with the conventional DM scattering, in which only the atomic nucleus is concerned and the electron contribution is negligible. We have described the SQUID-paramagnetic absorber experimental system that, having an energy resolution of 15 eV, is very suitable for the registration of DM particles and their interactions. Technical details about this system are discussed, and several modes of its operation are briefly mentioned.

In the context of "unifying" trend, clearly dominant in the modern elementary particle physics, we have also proposed a simple model, where we try to consider the corpuscular Dark Matter and non-corpuscular Dark Energy from uniform position. In this proposed model, the Dark Energy is an absolutely continuous substance, playing the role of vacuum for metastable excitations, which can be identified as Dark Matter particles. Estimates, carried out in the framework of this model, indicate the possibility of experimental detection of the "ether wind" pressure, created by the non-corpuscular incoming flow, corresponding to the galactic orbital motion of the Earth. It is argued that these types of investigations could be performed by using the SQUID-magnetostrictor experimental system.